\documentclass[floafix,prd,showpacs,showkeys,preprintnumbers,nofootinbib,superscriptaddress,11pt]{revtex4-1}
\usepackage[utf8]{inputenc}
\usepackage[sort&compress]{natbib}
\usepackage{ulem}
\usepackage{bm}
\usepackage{times}
\usepackage{amssymb,amsbsy,amsmath,amsfonts}
\usepackage{graphicx}
\usepackage{epstopdf}
\usepackage{float}
\usepackage{color}
\usepackage{morefloats}
\usepackage{rotating}
\usepackage{srcltx}
\usepackage{slashed}
\usepackage{subfigure}
\usepackage{multirow}
\usepackage{verbatim}
\usepackage{hyperref}
\usepackage{overpic}
\usepackage{tabularx}

\begin{document}

\title{Productions of  $D^*_{s0}(2317)$ and  $D_{s1}(2460)$  in  $B_{(s)}$ and  $\Lambda_b(\Xi_b)$   decays }

\author{Ming-Zhu Liu}
\affiliation{
Frontiers Science Center for Rare isotopes, Lanzhou University,
Lanzhou 730000, China}
\affiliation{ School of Nuclear Science and Technology, Lanzhou University, Lanzhou 730000, China}

\author{Xi-Zhe Ling}
\affiliation{Institute of High Energy Physics,
Chinese Academy of Sciences,
Beijing 100049, China}

\author{Li-Sheng Geng}\email{ lisheng.geng@buaa.edu.cn}
\affiliation{School of Physics, Beihang University, Beijing 102206, China}
\affiliation{Beijing Key Laboratory of Advanced Nuclear Materials and Physics, Beihang University, Beijing 102206, China}
\affiliation{Peng Huanwu Collaborative Center for Research and Education, Beihang University, Beijing 100191, China}
\affiliation{Southern Center for Nuclear-Science Theory (SCNT), Institute of Modern Physics, Chinese Academy of Sciences, Huizhou 516000, China}

\begin{abstract}

Recent studies show that $D_{s0}^{\ast}(2317)$ and $D_{s1}(2460)$ contain large molecular components.  In this work, we employ the naive factorization approach to calculate the production rates of $D_{s0}^{\ast}(2317)$ and $D_{s1}(2460)$ as hadronic molecules in $B_{(s)}$ and  $\Lambda_b(\Xi_b)$  decays, where their decay constants are estimated in the effective Lagrangian approach.  With the so-obtained decay constants $f_{D_{s0}^{\ast}(2317)}$ and $f_{D_{s1}(2460)}$, we calculate the branching fractions of the $b$-meson decays  $B_{(s)}\to \bar{D}_{(s)}^{(*)}D_{s0}^*$ and $B_{(s)}\to \bar{D}_{(s)}^{(*)}D_{s1}$ and the $b$-baryon decays  $\Lambda_b(\Xi_{b}) \to \Lambda_c(\Xi_{c}) D_{s0}^*$ and $\Lambda_b(\Xi_{b}) \to \Lambda_c(\Xi_c) D_{s1}$. Our results show that the production rates of  $D_{s0}^{\ast}(2317)$ and $D_{s1}(2460)$  in the $B_s$, $\Lambda_b$ and $\Xi_b$ decays are rather large that future experiments could observe them.    In particular,  we demonstrate that one can extract the decay constants of hadronic molecules via the triangle mechanism because of the equivalence of the triangle mechanism to the tree diagram established in calculating the decays  $B \to \bar{D}^{(*)}D_{s0}^{\ast}(2317)$ and   $B \to \bar{D}^{(*)}D_{s1}(2460)$.

\end{abstract}


\maketitle

\section{Introduction}

In 2003, the BaBar Collaboration discovered   $D_{s0}^*(2317)$ in the $D_{s}^+ \pi^0$ mass distribution in the $e^+ e^-$  annihilation process~\cite{BaBar:2003oey}, which was later confirmed by the CLEO Collaboration~\cite{CLEO:2003ggt} and   Belle Collaboration~\cite{Belle:2003kup} in the same process. 
Moreover,  the BESIII Collaboration observed the $D_{s0}^*(2317)$ in the process of $e^+e^- \to D_{s}^{*+} D_{s0}^{*-}(2317)$~\cite{BESIII:2017vdm}. In addition to the above inclusive processes, $D_{s0}^*(2317)$ was also observed in the exclusive process of the $B$ decay by the  Belle Collaboration~\cite{Belle:2003guh} and BaBar Collaboration~\cite{BaBar:2004yux}.      $D_{s1}(2460)$ as the heavy quark spin symmetry (HQSS) partner of $D_{s0}^*(2317)$  was first discovered in the $D_{s}^{\ast+} \pi^0$ mass distribution by the CLEO Collaboration~\cite{CLEO:2003ggt}, and then confirmed by several other experiments~\cite{Belle:2003kup,Belle:2003guh,BaBar:2003cdx}. Treating  $D_{s0}^*(2317)$ and  $D_{s1}(2460)$ as conventional $P$-wave  $c\bar{s}$  mesons, the masses obtained in the Goldfrey-Isgur (GI) model are larger than the experimental ones by 140 MeV and 100 MeV~\cite{Godfrey:1985xj}, which have motivated extensive discussions on their nature.

  By analyzing their masses, several interpretations were proposed for the internal structure of $D_{s0}^*(2317)$ and  $D_{s1}(2460)$.  In Ref.~\cite{Song:2015nia}, the authors found that the masses of  $D_{s0}^*(2317)$ and  $D_{s1}(2460)$ still deviate from the experimental data, even adding the screen potential to the conventional quark model. However,  as the $DK$ and $D^*K$ components were embodied into the conventional quark model, the mass puzzles of  $D_{s0}^*(2317)$ and  $D_{s1}(2460)$  are  esolved~\cite{Albaladejo:2018mhb,Yang:2021tvc, Luo:2021dvj}, which  indicate that the $D^{(*)}K$ components play an important role in forming $D_{s0}^*(2317)$ and  $D_{s1}(2460)$.    
  Therefore,  $D_{s0}^*(2317)$ and  $D_{s1}(2460)$  are proposed to be hadronic molecules of  $DK$ and $D^*K$ to explain their masses, especially their mass splitting~\cite{Barnes:2003dj, Gamermann:2006nm, Guo:2006fu,Xie:2010zza,Cleven:2010aw,Guo:2015dha,Wu:2019vsy}. It should be noted that in the lattice QCD simulation of the $DK$ interaction, a bound state below the $DK$ mass threshold was identified~\cite{Liu:2012zya,Mohler:2013rwa,Lang:2014yfa,Bali:2017pdv,Alexandrou:2019tmk}.  Furthermore, with the $D^{(*)}K$ potentials supplemented by the  $c\bar{s}$ core couplings to the $D^{(*)}K$ components, $D_{s0}^*(2317)$ and $D_{s1}(2460)$ can be dynamically generated~\cite{Weinberg:1965zz,MartinezTorres:2014kpc,Albaladejo:2016hae,Song:2022yvz},  indicating that the  $D^{(*)}K$  molecular components account for a large proportion of their wave functions in terms of the Weinberg compositeness rule $1-Z$~\cite{Weinberg:1965zz}.  Studying the masses of $D_{s0}^*(2317)$ and $D_{s1}(2460)$, one can conclude that they contain   both  $D^{(*)}K$   molecular components  and  {the $c\bar{s}$  cores}.   The next natural  step forward is to study their decays.

According to the review of particle physics (RPP)~\cite{ParticleDataGroup:2022pth}, the $D_{s0}^{*+}(2317)$  dominantly decays to $D_{s}^+\pi^0$, which means that  $D_{s0}^{*+}(2317)$ must be narrow since the decay of $D_{s0}^{*+}(2317) \to D_{s}^+\pi^0$ breaks isospin. The       
   dominant {decays}  of $D_{s1}^+(2460)$ into $D_{s}^{*+}\pi^0$ and $D_s^+\gamma$  { are }  responsible for its narrow width. The narrow widths of  $D_{s0}^*(2317)$ and $D_{s1}(2460)$ are quite different from the widths of their SU(3)-flavor partners $D^*_0(2300)$ and $D_1(2430)$, which {reflect}  the exotic properties of these excited charmed mesons.   In Refs.~\cite{Fayyazuddin:2003aa,Ishida:2003gu,Wei:2005ag,Nielsen:2005zr,Song:2015nia}, the authors proposed that the decays of $D_{s0}^*(2317) $ and   $D_{s1}(2460)$ as the $c\bar{s}$ excited states into $D_{s}\pi$ and  $D_{s}^{*}\pi$  proceed via the $\pi-\eta$ mixing, resulting in widths of  tens of keV. Treating    $D_{s0}^*(2317)$ and   $D_{s1}(2460)$ as hadronic molecules, 
 their widths are of the order of $100$~keV~\cite{Faessler:2007gv,Faessler:2007us,Fu:2021wde}. Up to now, there are no precise experimental measurements of the widths of  $D_{s0}^*(2317)$ and   $D_{s1}(2460)$, but  only their upper limits of $3.8$ MeV. From the perspective of their widths, one can obtain the same conclusion as from the studies of their masses regarding the nature of $D_{s0}^*(2317)$ and   $D_{s1}(2460)$. It is worth noting that a model-independent method has been proposed to verify the molecular nature of $D_{s0}^*(2317)$   by experimental searches for its three-body counterparts $DDK$ and $D\bar{D}K$~\cite{Wu:2019vsy,Huang:2019qmw,Wu:2022ftm,Wu:2022wgn}.   

The discoveries of  $D_{s0}^*(2317)$ and   $D_{s1}(2460)$  in the inclusive and exclusive processes in $e^+e^-$ collisions triggered a series of theoretical works to investigate their production mechanism.  Assuming $D_{s0}^*(2317) $ and   $D_{s1}(2460)$ as  $D^{(*)}K$ hadronic molecules and  $c\bar{s}$ excited states, Wu et al. estimated that their production rates in $e^+ e^-$ collisions are of the order of $10^{-3}$~\cite{Wu:2022wgn},   consistent with the experimental data~\cite{BaBar:2006eep}.   As for the exclusive process,   Faessler et al. calculated the decays $B\to D_{s0}^*(2317)\bar{D}^{(*)}$ and $B\to D_{s1}(2460)\bar{D}^{(*)}$  assuming   $D_{s0}^*(2317)$ and  $D_{s1}(2460)$ as $D^{(*)}K$ molecules~\cite{Faessler:2007cu}. The results are a bit smaller than the experimental data. Assuming  $D_{s0}^*(2317)$ and  $D_{s1}(2460)$  as  $c\bar{s}$  excited states, the decays $B\to D_{s0}^*(2317)\bar{D}^{(*)}$ and $B\to D_{s1}(2460)\bar{D}^{(*)}$ were investigated as well, but the results suffer from large uncertainties~\cite{Colangelo:1991ug,Veseli:1996yg,Cheng:2003sm,Colangelo:2005hv,Cheng:2006dm,Segovia:2012yh}. 
Moreover, the productions  of $D_{s0}^*(2317)$ and   $D_{s1}(2460)$    in the $\Lambda_b$ decays have been explored~\cite{Datta:2003yk}, finding that  their  production rates in the $\Lambda_b$ decays  are larger than those in the corresponding $B$ decays.  Recently, the $DK$ femtoscopic correlation function was investigated to elucidate the nature of   $D_{s0}^*(2317)$~\cite{Liu:2023uly,Ikeno:2023ojl}, which can be accessed in high energy nucleon-nucleon collisions in the future~\cite{STAR:2015kha,ALICE:2020mfd}.

Until now, the  $D_{s0}^*(2317) $ and   $D_{s1}(2460)$ have only been observed in the exclusive process  via  $B$ decays.  In this work, we systematically explore  the productions of $D_{s0}^*(2317)$ and   $D_{s1}(2460)$  in $B_{(s)}$ and  $\Lambda_b(\Xi_b)$  decays with the factorization ansatz~\cite{Chau:1982da,Chau:1987tk}. Following Ref.~\cite{Faessler:2007cu},  we employ the effective Lagrangian approach to estimate the decay constants of $ {D_{s0}^*(2317)}$ and $ {D_{s1}(2460)}$,  which are dynamically generated via the $DK-D_s\eta$ and $D^*K-D_s^*\eta$ coupled-channel potentials described by the contact-range effective field theory (EFT), and then calculate the production rates of $D_{s0}^*(2317) $ and   $D_{s1}(2460)$ in $B_{(s)}$ and  $\Lambda_b(\Xi_b)$  decays. Another motivation of this work is to test the universality of the approach that we proposed to calculate the decay constant of a hadronic molecule via the triangle mechanism~\cite{Wu:2023rrp}. Based on our previous study of the decays  $B\to D_{s0}^*(2317)\bar{D}^{(*)}$ and $B\to D_{s1}(2460)\bar{D}^{(*)}$ via the triangle mechanism, the decay constants of $D_{s0}^*(2317) $ and   $D_{s1}(2460)$ can be extracted~\cite{Liu:2022dmm}. The effective Lagrangian approach in this work can further check the validity of our approach\cite{Wu:2023rrp}.

This paper is organized as follows.  In Sec.~\ref{sec:formalism}, we introduce the effective Lagrangian approach to calculate the productions of  $D_{s0}^*(2317)$ and   $D_{s1}(2460)$  in $B_{(s)}$ and  $\Lambda_b(\Xi_b)$  decays and the decay constants of $f_{D_{s0}^*}$ and $f_{D_{s1}}$. Results and discussions are given in  Sec.~\ref{secresults}, followed by a summary in Sec.~\ref{sum}.

\section{ Theoretical formalism }
\label{sec:formalism}

\subsection{ Effective Lagrangians for nonleptonic Weak decays  }

\begin{figure}[ttt]
\centering
\includegraphics[width=0.76\columnwidth]{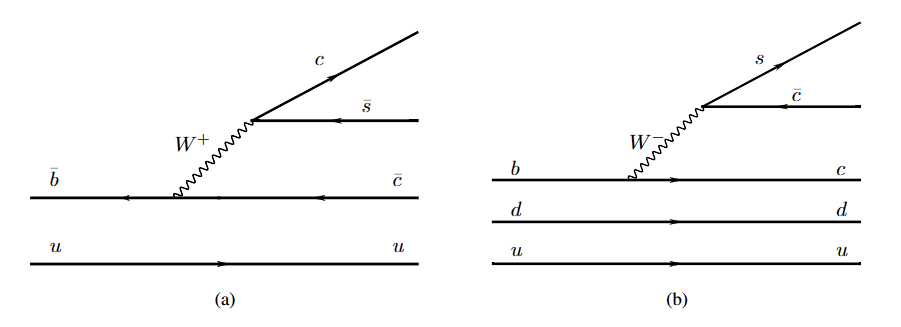}
\caption{\label{quarktopigy} External $W$-emission mechanism  for $B^+ \to \bar{D}^{(\ast) 0} D_{s}^{(\ast)+}$~(a) and  $\Lambda_b \to \Lambda_c D_{s}^{(\ast)-}$ (b).
}
\end{figure}

In this work,   we focus on  the productions of $D_{s0}^*(2317)$ and   $D_{s1}(2460)$  in $B_{(s)}$ and  $\Lambda_b(\Xi_b)$  decays. 
At quark level, the Cabibbo-favored  decays   $B^+ \to \bar{D}^{(\ast) 0} D_{s}^{(\ast)+}$ and  $\Lambda_b \to \Lambda_c D_{s}^{(\ast)-}$  mainly proceed via the external $W$-emission mechanism shown in Fig.~\ref{quarktopigy}  according to the topological  classification of weak decays~\cite{Chau:1987tk,Ali:1998eb,Ali:2007ff,Li:2012cfa}, which the naive factorization approach can well describe.  With the factorization ansatz~\cite{Bauer:1986bm},  the amplitudes of the weak decays  $B^+ \to \bar{D}^{(\ast) 0} D_{s}^{(\ast)+}$ and  $\Lambda_b \to \Lambda_c D_{s}^{(\ast)-}$   can be expressed as products of two current matrix elements
\begin{eqnarray}\label{Ds-KK}
\mathcal{A}\left(B^{+}\to D_{s}^{(\ast)+} \bar{D}^{(\ast) 0}\right)&=&\frac{G_{F}}{\sqrt{2}} V_{cb}V_{cs} a_{1}\left\langle D_{s}^{(\ast)+}|(s\bar{c})| 0\right\rangle\left\langle \bar{D}^{(\ast) 0}|(c \bar{b})| B^{+}\right\rangle, \\
\mathcal{A}\left(\Lambda_b \to D_{s}^{(\ast)-} \Lambda_c \right)&=&\frac{G_{F}}{\sqrt{2}} V_{cb}V_{cs} a_{1}\left\langle D_{s}^{(\ast)-}|(\bar{s} c)| 0\right\rangle\left\langle \Lambda_c|(\bar{c} b)| \Lambda_b\right\rangle, 
\end{eqnarray}
where  $G_F$ is the Fermi constant, $V_{cb}$ and $V_{cs}$ are  the Cabibbo-Kobayashi-Maskawa (CKM) matrix elements, and $a_{1}$ is the effective Wilson coefficient. In principle,  $a_{1}$ is expressed by the Wilson Coefficients of the QCD-corrected effective weak Hamiltonian, which depends on the renormalization scale~\cite{Chau:1987tk,Cheng:1993gf,Cheng:2010ry}.   In this work, we parameterize the non-factorization contributions with the effective Wilson coefficient $a_1$, which can be determined by reproducing relevant experimental data.

The current matrix elements of   $\left\langle \bar{D}^{ 0}|(c \bar{b})| B^{+}\right\rangle$ and $\left\langle \bar{D}^{ \ast0}|(c \bar{b})| B^{+}\right\rangle$ describing the hadronic transitions   are parameterized by six form factors~\cite{Cheng:2006dm} 
\begin{eqnarray}
\label{formfacotes}
&&\left\langle \bar{D}^{\ast0}|(c\bar{b})| B^{+}\right\rangle=\epsilon_{\alpha}^{*}\left\{-g^{\mu \alpha} (m_{\bar{D}^{\ast0}}+m_{B^{+}}) A_{1}\left(q^{2}\right)+P^{\mu} P^{\alpha} \frac{A_{2}\left(q^{2}\right)}{m_{\bar{D}^{\ast0}}+m_{B^{+}}}\right. +i \varepsilon^{\mu \alpha \beta \gamma}P_\beta q_\gamma  \\  \nonumber
&&\left.\frac{V\left(q^{2}\right)}{m_{\bar{D}^{\ast0}}+m_{B^{+}}} +q^{\mu} P^{\alpha} \left[\frac{m_{\bar{D}^{\ast0}}+m_{B^{+}}}{q^{2}}A_{1}\left(q^{2}\right)-\frac{m_{B^{+}}-m_{\bar{D}^{\ast0}}}{q^{2}}A_{2}\left(q^{2}\right)-\frac{2m_{\bar{D}^{\ast0}}}{q^{2}}A_{0}\left(q^{2}\right)\right]\right\},
\\  \nonumber
\end{eqnarray}
\begin{eqnarray}
 \left\langle \bar{D}^{0}|(c \bar{b} )| B^{+}\right\rangle =\left[P^{\mu}-\frac{m_{B^+}^2-m_{\bar{D}^0}^2}{q^2}q_{\mu}\right] F_{1}(q^2)+\frac{m_{B^+}^2-m_{\bar{D}^0}^2}{q^2}q_{\mu} F_{0}(q^2),   
\end{eqnarray}
where the momenta $q=p_{B^+}-p_{\bar{D}^{(*)0}}$ and $P=p_{B^+}+p_{\bar{D}^{(*)0}}$, and $F_{1}(q^2)$,  $F_{2}(q^2)$, $A_{0}(q^2)$, $A_{1}(q^2)$, $A_{2}(q^2)$, and $V(q^2)$  are form factors.   
The current matrix element $\langle \Lambda_c(p^{\prime})|\bar{c} b|\Lambda_b(p)\rangle$ is given by~\cite{Gutsche:2018utw} 
\begin{eqnarray}
\langle \Lambda_c(p^{\prime})|\bar{c} b|\Lambda_b(p)\rangle&=&
\bar{u}(p^{\prime})[f_{1}^V(q^2)\gamma_{\mu}-f_2^V(q^2)\frac{i\sigma_{\mu\nu}q^{\nu}}{
m}+f_3^V(q^2)\frac{q^{\mu}}{m}  \\ \nonumber
&& -(f_{1}^A(q^2)\gamma_{\mu}-f_2^A(q^2)\frac{i\sigma_{\mu\nu}q^{\nu}}{
m}+f_3^A(q^2)\frac{q^{\mu}}{m})\gamma^5
]u(p),
\end{eqnarray}
where $\sigma^{\mu\nu}=\frac{i}{2}(\gamma^\mu\gamma^\nu-\gamma^\nu\gamma^\mu)$ and $q=p-p^{\prime}$, and $f_{i}^{V/A}(q^2)$ are form factors.  In general, the form factors are  parameterized in the following form:
\begin{equation}
\frac{F_i(0)}{1-a\, \frac{q^2}{m^2}+ b\, (\frac{q^2}{m^2})^2}, 
\end{equation}
where $F_{i}(0)$, $a$, and $b$ are parameters determined in phenomenological models. In this work, we take these parameters determined in the quark model~\cite{Verma:2011yw,Gutsche:2015mxa,Faustov:2018ahb}.

The current matrix element $\left\langle D_{s}^{+}|(s\bar{c})| 0\right\rangle$ describes the process of creating a  $D_s^{+}$ meson from the vacuum via the axial current, which is parameterized by the decay constant $f_{D_{s}^{+}}$ and the momentum of the $D_{s}^+$ meson.  Following Ref.~\cite{Cheng:2003sm},   the current matrix elements for the $D_{s}$, $D_{s}^*$, $D_{s0}^*(2317)$, and $D_{s1}(2460)$ mesons created from the vacuum are 
\begin{eqnarray}
\label{decayconstant}
\left\langle D_{s}^{+}|(s  \bar{c})| 0\right\rangle & =&i f_{D_{s}^{+}} p^{\mu}_{D_{s}^{+}}, ~~~~~~
\left\langle D_{s}^{\ast+} |(s \bar{c} )| 0\right\rangle = m_{D_{s}^{\ast+}}f_{D_{s}^{\ast+}}\epsilon_\mu^*,  \\ \nonumber  ~~ \left\langle D_{s0}^{*+}|(s \bar{c})| 0\right\rangle& =& f_{D_{s0}^{*+}} p^{\mu}_{D_{s0}^{*+}}, ~~~~~
\left\langle D_{s1}^+ |(s \bar{c})| 0\right\rangle = m_{ D_{s1}^+}f_{ D_{s1}^+}\epsilon_\mu^*.  
\end{eqnarray}
 The  decay constants of $D_{s}$ and $D_{s}^*$  as $c\bar{s}$ ground states are determined to be $f_{D_{s}}=250$ MeV and   $f_{D_{s}^*}=272$ MeV~\cite{Verma:2011yw}. Due to the exotic properties of  $D_{s0}^*(2317)$ and $D_{s1}(2460)$,  the estimations of the decay constants $f_{D_{s0}^*}$ and $f_{D_{s1}}$ are quite uncertain.  In this work, we estimate the values of  $f_{D_{s0}^*}$ and $f_{D_{s1}}$ in the molecular picture.   In addition,  assuming SU(3)-flavor symmetry, the $B \to \bar{D}^{(*)}$ and $\Lambda_b \to \Lambda_c$ transitions can be related with the   $B_{s} \to \bar{D}_s^{(*)}$ and $\Xi_b \to \Xi_c$ transitions, and the production mechanism of   $D_{s0}^{*}(2317)$ and $D_{s1}(2460)$ in the $B_s$ and $\Xi_b$ decays are similar to those in the $B$ and $\Lambda_b$ decays as illustrated in Fig.~\ref{quarktopigy}. In the following, we only present the amplitudes for the decays $B\to \bar{D}^{(\ast)}D_{s}^{(\ast)}$  and $\Lambda_b\rightarrow\Lambda_c \bar{D}_s^{(*)}$,  and the amplitudes for the other decays have similar expressions.

With the above effective Lagrangian, we obtain the amplitudes for the decays  $B(k_0)\to \bar{D}^{(\ast)}(q_1)D_{s}^{(\ast)}(q_2)$: 
\begin{align}\label{am3}
\mathcal{A}(B\to D_{s}\bar{D}^{\ast})&= \frac{G_{F}}{\sqrt{2}}V_{cb}V_{cs} a_{1} f_{D_{s}}\{-q_{1}\cdot \varepsilon(q_{2})(m_{\bar{D}^{\ast}}+m_{B}) A_{1}\left(q_{1}^{2}\right)   \\ \nonumber 
&+(k_{0}+q_{2}) \cdot \varepsilon(q_{2}) q_{1}\cdot (k_{0}+q_{2}) \frac{A_{2}\left(q_{1}^{2}\right)}{m_{\bar{D}^{\ast}}+m_{B}} +(k_{0}+q_{2}) \cdot \varepsilon(q_{2}) \\ \nonumber       & 
[(m_{\bar{D}^{\ast}}+m_{B})A_{1}(q_{1}^2) -(m_{B}-m_{\bar{D}^{\ast}})A_{2}(q_1^2) -2m_{\bar{D}^{\ast}} A_{0}(q_{1}^2)]  \} , \\ \nonumber
\mathcal{A}(B\to D_{s}\bar{D})&=\frac{G_{F}}{\sqrt{2}}V_{cb}V_{cs} a_{1}^{\prime}f_{D_{s}}(m_{B}^2-m_D^2)F_{0}(q_{1}^2),
\\ \nonumber
\mathcal{A}(B\to D_{s}^{\ast}\bar{D})&= \frac{G_{F}}{\sqrt{2}}V_{cb}V_{cs} a_{1}^{*}m_{D_{s}^{\ast}}f_{D_{s}^{\ast}}(k_{0}+q_{2})\cdot \varepsilon(q_{1})F_{1}(q_{1}^2), \\ \nonumber
 \mathcal{A}(B\to D_{s}^{\ast}\bar{D}^{\ast})&= \frac{G_{F}}{\sqrt{2}}V_{cb}V_{cs} a_{1}^{*\prime}m_{D_{s}^{\ast}}f_{D_{s}^{\ast}}\varepsilon_{\mu}(q_1)\left[(-g^{\mu \alpha} (m_{\bar{D}^{\ast}}+m_{B}) A_{1}\left(q_1^{2}\right)\right.  \\ \nonumber &+ \left. (k_0+q_2)^{\mu} (k_0+q_2)^{\alpha} \frac{A_{2}\left(q_1^{2}\right)}{m_{\bar{D}^{\ast}}+m_{B}}  
+i \varepsilon^{\mu \alpha \beta \gamma}(k_0+q_2)_\beta q_{1\gamma} \frac{V\left(q_1^{2}\right)}{m_{\bar{D}^{\ast}}+m_{B}}\right]\varepsilon_{\alpha}(q_{2}).
\end{align}

In terms of the effective Lagrangian of the weak decays $\Lambda_b( p)\rightarrow\Lambda_c(p^{\prime}) \bar{D}_s^{(*)}(q)$~\cite{Cheng:1996cs}, the corresponding amplitudes are written as  
\begin{equation}\label{Eq:weakdecayV1}
\begin{split}
    \mathcal{A}(\Lambda_b \to \Lambda_c\bar{D}_s)&=\bar{u}(p^{\prime})(A+B \gamma_5) u(p),\\
     \mathcal{A}(\Lambda_b \to \Lambda_c\bar{D}_s^*)&=-i\bar{u}(p^{\prime})(A_1\gamma_\mu\gamma_5+A_2\frac{p_{2\mu}}{m}\gamma_5+B_1\gamma_\mu +B_2\frac{p_{2\mu}}{m})u(p) \varepsilon^{\mu}(q),
\end{split}
\end{equation}
where  $A_1$, $A_2$, $B_1$, $B_2$, $A$, and $B$  represent the transition form factors of $\Lambda_{b}$ to $\Lambda_{c}$:
\begin{equation}\label{Eq:weakdecayV2}
\begin{split}
    A&=\lambda f_{D_s}[(m-m_2)f_1^V+\frac{m_1^2}{m}f_3^V], \quad
    B=\lambda f_{D_s}[(m+m_2)f_1^A-\frac{m_1^2}{m}f_3^A], \\
    A_1&=-\lambda f_{D_s^*}m_1[f_1^A-f_2^A\frac{m-m_2}{m}], \quad\quad
    B_1=\lambda f_{D_s^*}m_1[f_1^V+f_2^V\frac{m+m_2}{m}], \\
    A_2&=2\lambda f_{D_s^*}m_1f_2^A, \quad\quad\quad\quad\quad\quad\quad\quad
    B_2=-2\lambda f_{D_s^*}m_1f_2^V,
\end{split}
\end{equation}
with $\lambda=\frac{G_F}{\sqrt{2}}V_{cb}V_{cs}a_1$ and $m, m_1, m_2$ referring  to the masses of $\Lambda_b$, $\bar{D}_s^{(*)}$, and $\Lambda_c$.  

 With  the amplitudes  for the weak decays   given above, 
 one can compute the corresponding partial decay widths 
 \begin{eqnarray}
\Gamma=\frac{1}{2J+1}\frac{1}{8\pi}\frac{|\vec{p}|}{{ M}^2}{|\overline{{\mathcal{A}}}|}^{2},
\end{eqnarray}
{ where $J$ and $M$ are the total angular momentum and the mass of the initial state, $|\vec{p}|$ is the momentum of either final state in the rest frame of the initial state, $\mathcal{A}$ is the amplitude of the weak decay, and the overline indicates the sum over the polarizations of final states. }     

\subsection{ Decay Constants }

\begin{figure}[!h]
\centering
\includegraphics[width=0.75\columnwidth]{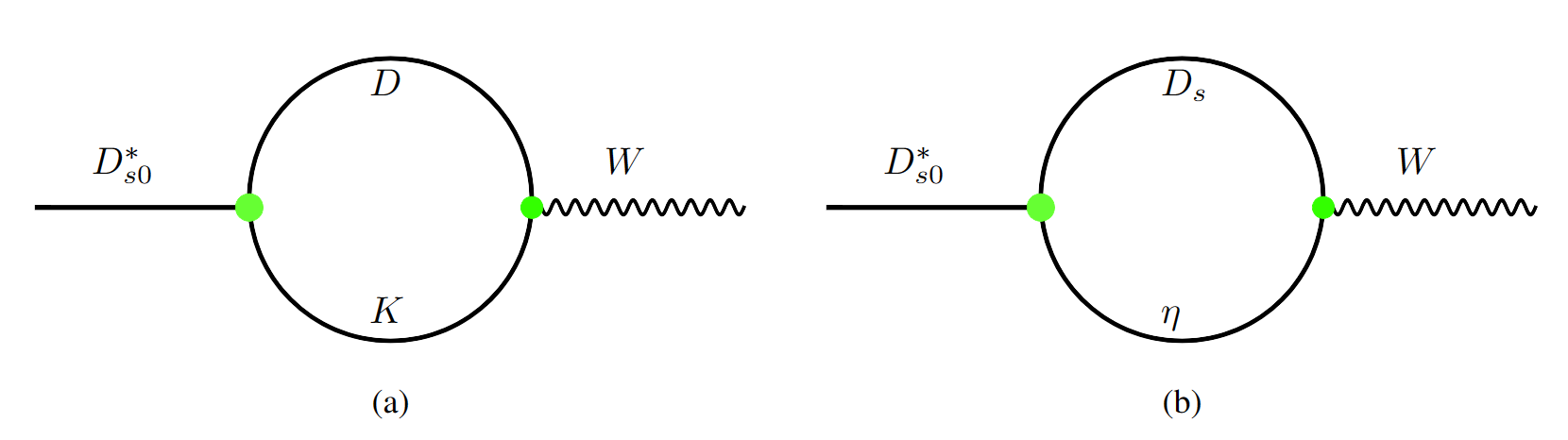}
\caption{\label{creating2317} Feynman diagrams for the $W$ boson transiting to  $D_{s0}^*(2317)$ in the $DK$ and $D_s\eta$ molecular picture. 
}
\end{figure}

The decay constants $f_{D_{s0}^*}$ and $f_{D_{s1}}$  are defined in Eq.(\ref{decayconstant}). To obtain the values of  $f_{D_{s0}^*}$ and $f_{D_{s1}}$, one usually constructs the amplitudes for the $D_{s0}^*$ and $D_{s1}$  created from the vacuum and then extracts the coefficients of $p^{\mu}_{D_{s0}^{*+}}$  and  $m_{ D_{s1}^+}\epsilon_\mu^*$~\cite{Colangelo:1991ug,Cheng:2003sm,Thomas:2005bu,Colangelo:2005hv}. Following the same principle, we calculate the $f_{D_{s0}^*}$ and $f_{D_{s1}}$  decay constants in the molecular picture.    Assuming that $D_{s0}^*$ is dynamically generated by the $DK$ and $D_s\eta$ coupled-channel interactions, the current matrix element $\left\langle D_{s0}^{*+}|(s\bar{c})| 0\right\rangle$ is illustrated in Fig.~\ref{creating2317}. Considering HQSS, we replace the above $D$ and $D_s$ mesons  with the $D^*$ and $D_s^*$ mesons, dynamically generating the $D_{s1}$.    In the following, we introduce the effective Lagrangian approach to calculate the decay constants of molecules.

The effective Lagrangians for  the  $D^{(\ast)}(D_{s}^{(\ast)})$ mesons transiting to the $K(\eta)$ mesons and  $W$ boson  are given by
\begin{eqnarray}
\mathcal{L}_{VDK}&=&f_{1}^{DK}(0)V^{\mu}(D\partial_{\mu}K-\partial_{\mu}DK),   ~~~ \mathcal{L}_{VD_s\eta}=f_{1}^{D_s\eta}(0)V^{\mu}(D_s\partial_{\mu}\eta-\partial_{\mu}D_s\eta)  \\ \nonumber
\mathcal{L}_{AD^*K}&=&(m_{D}+m_{K^*})A_{1}^{DK^*}(0)A^{\mu}D^*_{\mu}K,  ~~~\mathcal{L}_{AD_s^*\eta}=(m_{D_s}+m_{\phi})A_{1}^{D_s\phi}(0)A^{\mu}D^*_{s\mu}\eta
\end{eqnarray}
where  $f_{1}^{DK}(0)[f_{1}^{D_s\eta}(0)]$ and $A_{1}^{DK^*}(0)[A_{1}^{D_s\phi}(0)]$ are the form factors at $q^2=0$. Such parameters can be determined by fitting  the corresponding semileptonic branching fractions. In this work, we take the following values: $f_0^{DK}=0.74$~\cite{Cheng:2010ry,Cheng:2016ejf,Zhang:2018jtm},  $A_1^{DK^*}=0.78$, $A_1^{D_s\phi}=0.77$~\cite{Chang:2019mmh}, $F_{+}^{DK}=0.77$, $F_{+}^{D_s\eta}=0.49$, $A_{0}^{DK^*}=2.08$, $A_{0}^{D_s\phi}=2.13$
~\cite{Ivanov:2019nqd}

The effective Lagrangians describing the couplings of the hadronic molecules to their constituents are written as 
\begin{eqnarray}
\mathcal{L}_{D_{s0}^*DK}&=&g_{D_{s0}^*DK}D_{s0}^*  DK,    ~~~~ \mathcal{L}_{D_{s0}^*D_s\eta}=g_{D_{s0}^*D_s\eta}D_{s0}^*D_s\eta \\ \nonumber
\mathcal{L}_{D_{s1}D^*K}&=&g_{D_{s1}D^*K}D_{s1}^{\mu}  D^*_{\mu}K,   ~~ \mathcal{L}_{D_{s1}D_s^*\eta}=g_{D_{s1}D_s^*\eta}D_{s1}^{\mu}  D_{s\mu}^{*}K,
\end{eqnarray}
where $g_{D_{s0}^*DK}$, $g_{D_{s0}^*D_s\eta}$, $g_{D_{s1}D^*K}$, and $g_{D_{s1}D_s^*\eta}$ are the coupling constants between   $D_{s0}^*(D_{s1})$ and their constituents. In this work, we employ the contact-range EFT to dynamically generate the   $D_{s0}^*$ and $D_{s1}$ and further determine the 
 couplings between the molecular states and their constituents from the residues of the corresponding poles, which are widely applied to study hadronic molecules~\cite{Xiao:2019aya,Du:2019pij,Pan:2023hrk}.

With the above preparations, we can write the amplitude of Fig.~\ref{creating2317} as
\begin{eqnarray}
\label{ds2317int}
\mathcal{A}_{a}&=&g_{D_{s0}^*DK} f_{1}^{DK}(0)\int \frac{d^4 q}{(2\pi)^4}\frac{1}{k_{1}^2-m_{1}^2}\frac{1}{k_{2}^2-m_{2}^2}(k_{1}^{\mu}-k_{2}^{\mu})\varepsilon_{\mu}(V),
\\ \nonumber
\mathcal{A}_{b}&=&g_{D_{s0}^*D_s\eta } f_{1}^{D_s\eta }(0)\int \frac{d^4 q}{(2\pi)^4}\frac{1}{k_{1}^2-m_{1}^2}\frac{1}{k_{2}^2-m_{2}^2}(k_{1}^{\mu}-k_{2}^{\mu})\varepsilon_{\mu}(V),
\end{eqnarray}
where the   subscripts 1 and 2 denote the  $D$ and $K$ mesons in amplitude $\mathcal{A}_{a}$,  and the $D_s$ and $\eta$ mesons in amplitude $\mathcal{A}_{b}$. Similarly, we obtain the amplitudes describing the  $D_{s1}$ created from the vacuum  as 
\begin{eqnarray}
\label{ds2460int}
\mathcal{A}_{a}&=&g_{D_{s1}D^*K}(m_{D}+m_{K^*})A_{1}^{DK^*}(0)\int \frac{d^4 q}{(2\pi)^4}\varepsilon_{\mu}(k_0)\frac{-g^{\mu\nu}+\frac{k_1^{\mu} k_1^{\nu}}{m_1^2}}{k_{1}^2-m_{1}^2}\frac{1}{k_{2}^2-m_{2}^2}\varepsilon_{\nu}(A),
\\ \nonumber
\mathcal{A}_{b}&=&g_{D_{s1}D_s^*\eta}(m_{D_s}+m_{\phi})A_{1}^{D_s\phi}(0)\int \frac{d^4 q}{(2\pi)^4}\varepsilon_{\mu}(k_0)\frac{-g^{\mu\nu}+\frac{k_1^{\mu} k_1^{\nu}}{m_1^2}}{k_{1}^2-m_{1}^2}\frac{1}{k_{2}^2-m_{2}^2}\varepsilon_{\nu}(A).
\end{eqnarray}
Once the amplitudes of Fig.~\ref{creating2317} are obtained, the decay constants $f_{D_{s0}^*}$ and $f_{D_{s1}}$ can be easily extracted considering their definitions. In the following, we show how to calculate the relevant loop functions in the dimensional regularisation scheme.

With the Feynman parameter approach, we obtain the following integrals  
\begin{eqnarray}
\int \frac{d^4k_1}{(2\pi)^4} \frac{1}{(k_1^2-m_1^2)} &=&\frac{m_1^2}{16\pi^2} (ln\frac{m_1^2}{\mu^2}-1),   \\ \nonumber
\int \frac{d^4k_1}{(2\pi)^4} \frac{1}{(k_1^2-m_1^2)[(p-k_1)^2-m_2^2]} &=&\frac{1}{16\pi^2}\int_{0}^{1}dx  ln\frac{\Delta^2}{\mu^2},  \\ \nonumber
\int \frac{d^4k}{(2\pi)^4}  \frac{k^2}{(k_1^2-m_1^2)[(p-k_1)^2-m_2^2]}  &=& \frac{1}{16\pi^2} \int_{0}^{1}dx [\Delta^2(2(ln\frac{\Delta^2}{\mu^2}-1)+1)+ p^2x^2ln\frac{\Delta^2}{\mu^2}], \\ \nonumber
\int \frac{d^4k}{(2\pi)^4}  \frac{k_1^{\mu}k_1^{\nu}}{(k_1^2-m_1^2)[(p-k_1)^2-m_2^2]}  &=& \frac{1}{16\pi^2} \int_{0}^{1}dx [ g^{\mu\nu}\frac{\Delta^2}{2}\,(ln\frac{\Delta^2}{\mu^2}-1)+ p^{\mu}p^{\nu}x^2\,ln\frac{\Delta^2}{\mu^2}],
\end{eqnarray}
where $\Delta^2=p^2x^2-m_{1}^2(x-1)-x(p^2-m_2^2)$, $p=k_1+k_2$,  and the renormalization scale $\mu$ depends on the specific  physical process under consideration. To extract the decay constants of $D_{s0}^*(2317)$ and $D_{s1}(2460)$,     the  loop functions of Eq.~(\ref{ds2317int}) and Eq.~(\ref{ds2460int})      are converted   into the following form
\begin{eqnarray}
\label{decayconstantv}
&&\int \frac{d^4k_1}{(2\pi)^4} \frac{k_1^{\mu}-k_2^{\mu}}{(k_1^2-m_1^2)[(p-k_1)^2-m_2^2]} =  \frac{p^{\mu}}{16\pi^2}\int_{0}^{1}dx (2x-1) ln\frac{\Delta^2}{\mu^2},   \\ \nonumber
&& \int \frac{d^4k_1}{(2\pi)^4} \frac{-g^{\mu\nu}+\frac{k_1^{\mu}k_1^{\nu}}{m_1^2}}{(k_1^2-m_1^2)[(p-k_1)^2-m_2^2]} =    -\frac{1}{16\pi^2}\int_{0}^{1}dx \{g^{\mu\nu}\left[ ln\frac{\Delta^2}{\mu^2} 
 - \frac{1}{2m_1^2} \Delta^2(ln\frac{\Delta^2}{\mu^2}-1)\right]  \\ \nonumber  && ~~~~~~~~~~~~~~~~~~~~~~~~~~~~~~~~~~~~~~~~~~~~~~~~~~~~~~~~~~~~~~~~~ -\frac{1}{m_1^2}p^{\mu}p^{\nu} x^2ln\frac{\Delta^2}{\mu^2} \}. 
\end{eqnarray}
Finally, we obtain the analytic form of the decay constants of  $D_{s0}^*(2317)$ and $D_{s1}(2460)$ 
\begin{eqnarray}
\label{fd2317coupling}
 f_{D_{s0}^*}^{m_1 m_2} &=&  g_{D_{s0}^* m_1 m_2} f_{1}^{ m_1 m_2}(0)\frac{1}{16\pi^2}\int_{0}^{1}dx (2x-1) ln\frac{\Delta^2}{\mu^2},   \\ 
 \label{fd2460coupling}
f_{D_{s1}}^{m_1 m_2} &=& \frac{g_{D_{s1} m_1 m_2}(m_{1}+m_{2})A_{1}^{m_1 m_2}(0)}{m_{D_{s1}}}\frac{1}{16\pi^2}\int_{0}^{1}dx ~ \left[ \frac{1}{2m_1^2} \Delta^2(ln\frac{\Delta^2}{\mu^2}-1) -ln\frac{\Delta^2}{\mu^2} \right],
\end{eqnarray} 
where $m_1$ and $m_2$ refer to  $D(D_s)$ and $K(\eta)$ for $D_{s0}^*(2317)$ and $D^*(D_s^*)$ and $K(\eta)$ for $D_{s1}(2460)$. The decay constants of  $D_{s0}^*(2317)$ and $D_{s1}(2460)$ are calculated as the sum of  $ f_{D_{s0}^*}^{DK}$ and $ f_{D_{s0}^*}^{D_s\eta}$
and the sum of $ f_{D_{s1}}^{D^*K}$ and $ f_{D_{s1}}^{D_s^*\eta}$.

\subsection{Contact-range effective field theory approach }

In the following, we briefly introduce the contact-range effective field theory (EFT) approach. The scattering amplitude $T$ is responsible for  the dynamical generations of  molecules, which is obtained by solving the following Lippmann-Schwinger equation
\begin{eqnarray}
T(\sqrt{s})=(1-VG(\sqrt{s}))^{-1}V,
\label{lsequation}
\end{eqnarray}
where $V$ is the coupled-channel potential determined by the contact-range EFT approach, and $G(\sqrt{s})$  is the loop function of the two-body propagator.  

The  coupled-channel potentials $V$ in matrix form  read
\begin{equation}
    V_{DK-D_s\eta}^{J^P=0^+}=\begin{pmatrix} 
     -2 C_a&  \sqrt{3} C_a\\ \sqrt{3} C_a& 0\end{pmatrix}, \quad
    V_{D^*K-D_s^*\eta}^{J^P=1^+}=\varepsilon(k_1)\cdot \varepsilon(k_1^{\prime}) \begin{pmatrix} 
     -2 C_a&  \sqrt{3} C_a\\ \sqrt{3} C_a& 0\end{pmatrix},
\end{equation}
where the coefficient $C_a$ needs to be determined by fitting the  $D_{s0}^*$ and $D_{s1}$ masses.    The loop functions of $D_{s0}^*$ and $D_{s1}$   are    
\begin{eqnarray}
\label{loopfunctionv}
G(\sqrt{s})^{D_{s0}^*} &=&\frac{1}{16\pi^2}\int_{0}^{1}dx  ln\frac{\Delta^2}{\mu^2},   \\ \nonumber
G(\sqrt{s})^{D_{s1}} &=& \frac{1}{16\pi^2}\int_{0}^{1}dx ~ \left[ ln\frac{\Delta^2}{\mu^2} - \frac{1}{2m_1^2} \Delta^2(ln\frac{\Delta^2}{\mu^2}-1)\right] 
\end{eqnarray}
with $\Delta^2=s x^2-m_{1}^2(x-1)-x(s-m_2^2)$. We note that the loop function of $D_{s1}$ contains an additional term, which is induced by the  $\frac{k_1^{\mu}k_1^{\nu}}{m_1^2}$ term in the loop integral.  One can see that the loop integrals depend on the renormalization scale $\mu$.  

With the potentials obtained above, we can search for poles generated by the coupled-channel interactions and  determine the 
 couplings between the molecular states and their constituents from the residues of the corresponding poles, 
\begin{eqnarray}
g_{i}g_{j}=\lim_{\sqrt{s}\to \sqrt{s_0}}\left(\sqrt{s}-\sqrt{s_0}\right)T_{ij}(\sqrt{s}),
\end{eqnarray}
where $g_{i}$ denotes the coupling of channel $i$ to the  dynamically generated state and $\sqrt{s_0}$ is the pole position.

\section {Results and Discussions}
\label{secresults}

\begin{table}[!h]
\caption{Masses and quantum numbers of relevant hadrons needed in this work~\cite{ParticleDataGroup:2022pth}. \label{mass}}
\begin{tabular}{ccc|ccc|ccc}
  \hline\hline
   Hadron & $I (J^P)$ & M (MeV) &    Hadron & $I (J^P)$ & M (MeV) &    Hadron & $I (J^P)$ & M (MeV)     \\
  \hline
     $K^{\pm}$ & $1/2(0^-)$ & $493.677$ & 
      $K^{0}$ & $1/2(0^-)$ & $497.611$ &  $\Lambda_b$ & $0(1/2^+)$ & $5619.60$ \\
   $\bar{D}^{0}$ & $1/2(0^-)$ & $1864.84$  &    $D^{-}$ & $1/2(0^-)$ & $1869.66$   & 
      $\Lambda_{c}^{+}$ & $0(1/2^+)$ & $2286.46$   \\
  $\bar{D}^{\ast0}$ & $1/2(1^-)$ & $2006.85$ &  $D^{\ast-}$ & $1/2(1^-)$ & $2010.26$   & 
      $B^{+}$ & $1/2(0^-)$ & $5279.34$ \\
  $J/\psi$ & $0(1^-)$ & $3096.90$& 
  $\eta_{c}$ & $0(0^-)$ & $2983.90$  & 
      $B^{0}$ & $1/2(0^-)$ & $5279.65$   \\
    $D_s^{\pm}$ & $0(0^-)$ & $1968.35$& 
$D_s^{\ast\pm}$ & $0(1^-)$ & $2112.2$  & 
      $B_s^{0}$ & $0(0^-)$ & $5366.91$    \\
 \hline \hline
\end{tabular}
\label{tab:masses}
\end{table}

In Table~\ref{tab:masses}, we tabulate the masses and quantum numbers of relevant particles.  One can see that there exists an unknown parameter $\mu$~(renormalization scale) in both Eq.~(\ref{decayconstantv}) and  Eq.~(\ref{loopfunctionv}), for which a consistent value is adopted in this work. First, we employ the contact-range EFT approach to dynamically generate the poles corresponding to $D_{s0}^*(2317)$ and $D_{s1}(2460)$  by varying  $\mu$ and then obtain the  $D_{s0}^*(2317)$ and $D_{s1}(2460)$  couplings to their constituents as well as their decay constants. With the so-obtained decay constants $f_{D_{s0^*}(2317)}$ and $f_{D_{s1}(2460)}$  we 
further study the productions of $D_{s0}^*(2317)$ and $D_{s1}(2460)$ in the $B_{(s)}$ and  $\Lambda_b(\Xi_b)$  decays, where the naive factorization approach works well as mentioned above.  In this work, assuming that the decay mechanisms of $ H_b \to H_c D_{s}^{(\ast)}$($H_b$ and $H_c$ denote bottom and charm hadrons of interest) and $H_b \to H_c D_{s0}^*(D_{s1})$ are the same\footnote{{  The productions of the   $D_{s}^{(*)}$ mesons in the $b$-favored  decays mainly occur via short-distance interactions, while those of the $D_{s0}^*(D_{s1})$ mesons mainly occur via long-distance interactions due to the exotic properties of the $D_{s0}^*(D_{s1})$ mesons.  The effects of long-range interactions in these decays are induced by final-state interactions via triangle diagrams~\cite{Cheng:2004ru,Yu:2017zst,Cao:2023csx}.   In this work, the long-distance effects are embodied into the decay constants of  $D_{s0}^*(D_{s1})$ mesons, implying that the production mechanisms of the $D_{s0}^*(D_{s1})$ mesons in $b$-flavored decays are similar to those of the $D_s^{(*)}$ mesons.    
}}, we parameterize the unknown non-factorization contributions with the effective Wilson coefficients.  In other words, we determine  $a_1$ by reproducing the experimental branching fractions of $ H_b \to H_c D_{s}^{(\ast)}$ decays, and then calculate the branching fractions of the corresponding decays $H_b \to H_c D_{s0}^*(D_{s1})$ using the so-obtained $a_1$.

 \begin{table}[!h]
 \centering
 \caption{$D_{s0}^*(2317)$ and $D_{s1}(2460)$ couplings to their constituents (in units of GeV).  \label{moleculecouplingsc} }
 \begin{tabular}{cccccccc}
 \hline\hline
 Couplings ~~~  &    $\mu=1.00$~~~ & $\mu=1.50$~~~ &  $\mu=2.00$~~~ & Fu et al.~\cite{Fu:2021wde}       \\
 \hline
 $g_{D_{s0}^*DK}$~~~   &11.75~~~& 11.92~~~  & 11.95~~~ & 9.4~~~\\
 $g_{D_{s0}^*D_s\eta }$~~~   &8.13~~~ &7.47~~~  & 7.32~~~ &7.4~~~  \\
 \hline
 $g_{D_{s1}D^*K}$~~~   & 12.06~~~& 12.16~~~  & 12.15~~~ & 10.1~~~\\
 $g_{D_{s1}D_s^*\eta }$~~~   &8.78~~~ &7.76~~~  & 7.53~~~ &7.9~~~  \\
\hline\hline
 \end{tabular}
 \end{table}

In Refs.~\cite{Gamermann:2006nm,Molina:2010tx,Chen:2023fgl}, the loop function is regularised in the dimensional regularization scheme, which shows that  $\mu$ is around $1.5$~GeV in the charm sector. To quantify the uncertainty of the renormalization scale, we vary $\mu$ from $1$~GeV to $2$~GeV in this work.  For $\mu$ of $1.0$~GeV,  $1.5$~GeV, and $2.0$~GeV, the values of $C_a$ for  $DK-D_s\eta$ ($D^*K-D_s^*\eta$)  contact-range potentials are  determined as $74.78$, $34.53$, and $25.04$ ($98.60$, $42.45$, and $30.34$).  With the so-obtained scattering amplitude $T$, we obtain the $D_{s0}^*(2317)$ and $D_{s1}(2460)$  couplings to their constituents shown in Table~\ref{moleculecouplingsc},  a bit different from the estimations of  Ref.~\cite{Fu:2021wde}\footnote{ One should note that an additional parameter, e.g.,  subtraction constant,  is introduced in Ref.~\cite{Fu:2021wde}.  }. Finally, the  decays constants of $D_{s0}^*(2317)$ and $D_{s1}(2460)$ are    determined  as shown    in Table~\ref{Tab:FormFactor1}.   One can see that  $f_{D_{s0}^*(2317)}$ is almost independent of the renormalization scale, as can also seen from Eq.~(\ref{fd2317coupling}).  The slight variation of $f_{D_{s0}^*(2317)}$ stems from  the weak dependence of the couplings  $g_{D_{s0}^*DK}$ and  $g_{D_{s0}^*D_s \eta}$ on the renormalization scale $\mu$. However,  the  
 decay constant $f_{D_{s1}(2460)}$ is dependent on  $\mu$ as shown in Table~\ref{Tab:FormFactor1}.        In the following calculations, we adopt the values of $f_{D_{s0^*}(2317)}$ and $f_{D_{s1}(2460)}$ at $\mu=1.5$~GeV, e.g., $f_{D_{s0}^*}=58.74$~MeV and  $f_{D_{s1}}=133.76$~MeV, which are  consistent with the results  of  Ref.~\cite{Faessler:2007cu}, but smaller than the results of lattice QCD~\cite{Bali:2017pdv}.

  \begin{table}[ttt]
 \centering
 \caption{Decay constants of $D_{s0}^*(2317)$ and $D_{s1}(2460)$ as  hadronic molecules (in units of MeV). \label{Tab:FormFactor1} }
 \begin{tabular}{cccccccc}
 \hline\hline
 Decay Constants ~~~  & $\mu=1000$~~~ & $\mu=1500$~~~ &  $\mu=2000$~~~ & Faessler et al.~\cite{Faessler:2007cu}       \\
 \hline
 $f_{D_{s0}^*(2317)}$~~~   & 59.36~~~& 58.74~~~  & 58.59~~~   & 67.1~~~\\
  $f_{D_{s1}(2460)}$~~~   & 56.10~~~ &133.76~~~  & 187.48~~~  & 144.5~~~  \\
\hline\hline
 \end{tabular}
 \end{table}

 \begin{table}[!h]
 \centering
 \caption{Values of  $F(0)^{B\to D^{(*)}}$, $a^{B\to D^{(*)}}$, $b^{B\to D^{(*)}}$ in the $B \rightarrow D^{(*)}$  transition  form factors and $F(0)^{B_s\to D_s^{(*)}}$, $a^{B_s\to D_s^{(*)}}$, $b^{B_s\to D_s^{(*)}}$  in the ${B_s\to D_s^{(*)}}$  transition  form factors~\cite{Verma:2011yw}. \label{BtoDformfactor} }
 \begin{tabular}{ccccccc|ccccccc}
 \hline\hline
   & $F_0$~~~ & $F_1$~~~ & $V$~~~ & $A_0$~~~ &  $A_1$~~~ & $A_2$~~~    &    &$F_0$~~~ & $F_1$~~~ & $V$~~~ & $A_0$~~~ &  $A_1$~~~ & $A_2$~~~\\
 \hline
 $F(0)^{B\to D^{(*)}}$&  0.67~~~  & 0.67~~~ & 0.77~~~ & 0.68~~~ & 0.65~~~ & 0.61~~~ &  $F(0)^{B_s\to D_s^{(*)}}$~~~ &  0.67~~~  & 0.67~~~ & 0.75~~~ & 0.66~~~ & 0.62~~~ & 0.57~~~\\
 $a^{B\to D^{(*)}}$  & 0.63~~~ &  1.22~~~  & 1.25~~~ & 1.21~~~ & 0.60~~~ & 1.12~~~  &  $a^{B_s\to D_s^{(*)}}$~~~  & 0.69~~~ &  1.28~~~  & 1.37~~~ & 1.33~~~ & 0.76~~~ & 1.25~~~\\
 $b^{B\to D^{(*)}}$ & -0.01~~~  & 0.36~~~  & 0.38~~~ &0.36~~~ & 0.00~~~ & 0.31~~~ &  $b^{B_s\to D_s^{(*)}}$~~~ & 0.07~~~  & 0.52~~~  & 0.67~~~ &0.63~~~ & 0.13~~~ & 0.56~~~\\
\hline\hline
 \end{tabular}
 \end{table}

\begin{figure}[ttt]
\centering
\includegraphics[width=0.75\columnwidth]{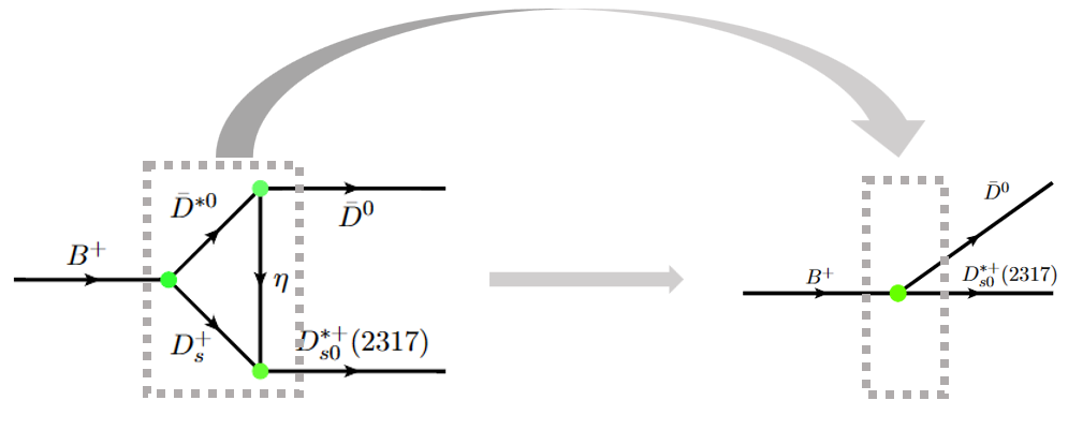}
\caption{\label{rtis} Equivalence of the
triangle diagram and the tree diagram depict the decay of $B^{+}\to \bar{D}^0 D_{s0}^{*+}(2317)$.  
}
\end{figure}

Up to now, $D_{s0}^*(2317)$ and $D_{s1}(2460)$ have only been observed in $B$ decays. Therefore, we first focus on the decays of $B\to \bar{D}^{(\ast)}D_{s}^{(\ast)}$ and $B\to \bar{D}^{(\ast)}D_{s0}^*(D_{s1})$.   In Table~\ref{BtoDformfactor}, we present the parameters of the form factors in the $B \to D$ and $B \to D^*$ transitions, which are taken from  Ref.~\cite{Verma:2011yw}. By reproducing the experimental branching fractions of the decays $B\to \bar{D}^{(\ast)}D_{s}^{(\ast)}$, we determine the   effective Wilson coefficient $a_1$.    Using   $a_1$   and   $f_{D_{s0}^*}(f_{D_{s}})$ obtained above,   we calculate the branching fractions of the decays $B\to \bar{D}^{(\ast)}D_{s0}(D_{s1})$ , which are shown in Table~\ref{resultsbtodds}. Our results are a bit smaller than those of Ref.~\cite{Faessler:2007cu} because of the smaller values for the decay constants and the effective Wilson coefficient. Interestingly,  our results are consistent with our previous calculations using the triangle mechanism except for the decay $B^+ \to \bar{D}^{\ast 0}D_{s0}^{*+}$~\cite{Liu:2022dmm} \footnote{We note that in the triangle diagram, the relative phase between the $D^{(*)}$ exchange and the $\eta$ exchange is fixed in such a way that they add constructively, which produces results in better agreement with data. This may not hold in the decay of $B^+ \to \bar{D}^{\ast 0}D_{s0}^{*+}$.     }, 
which {indicate} that the triangle diagram and tree diagram accounting for the decays  $B\to \bar{D}^{(\ast)}D_{s0}(D_{s1})$  are equivalent. In principle, one can replace the triangle diagram with one vertex, resulting in an effective description for the weak decay $B^{+}\to \bar{D}^0 D_{s0}^{*+}(2317)$ at tree level as shown in Fig.~\ref{rtis}, which indicates that it is reasonable to extract the decay constants of hadronic molecules using the triangle mechanism. In Ref.~\cite{Wu:2023rrp}, with this approach, we extract the decay constants of $X(3872)$ as a $\bar{D}D^*$  molecule. Here, we note that the relative phase among various amplitudes may lead to uncertainties in extracting the decay constants. As a result, it is better to select relevant amplitude with no or small relative phases.

\begin{table}[!h]
\centering
\caption{ {  Branching fractions   ($10^{-3}$) of $B\to \bar{D}^{(\ast)}D_{s0}^{*}(D_{s1})$.  } \label{resultsbtodds}
}
\begin{tabular}{c c c | c c c c c c c}
  \hline \hline
   Decay modes    &~~~~ Exp~\cite{ParticleDataGroup:2022pth}  &  ~~$a_1$~~      &   Decay modes    &~~~~ $\mathrm{Ours}$   & ~~~~ Triangle~\cite{Liu:2022dmm}  & ~~~~ Exp~\cite{ParticleDataGroup:2022pth}
         \\ \hline 
      $B^{+} \to \bar{D}^{0}D_{s}^{+}$    & ~~~~ $9.0\pm 0.9$    &   ~~0.80~~    &       $B^{+} \to \bar{D}^{0}D_{s0}^{*+}(2317)$    & ~~~~ $0.48$    & ~~~~        $0.68$    & ~~~~        $0.80^{+0.16}_{-0.13}$
        \\
                           $B^{+} \to \bar{D}^{\ast0}D_{s}^{+} $ & ~~~~ $8.2\pm 1.7$ & ~~0.93~~   &                    $B^{+} \to \bar{D}^{\ast0}D_{s0}^{*+}(2317) $ & ~~~~ $0.39$    & ~~~~ $1.21$   & ~~~~      $0.90^{+0.70}_{-0.70}$ 
         \\
                 $B^{+} \to \bar{D}^{0}D_{s}^{\ast+} $ & ~~~~ $7.6\pm 1.6$ &     ~~0.81~~     &    $B^{+} \to \bar{D}^{0}D_{s1}^{+}(2460) $ & ~~~~ $1.39$ &     ~~~~    $1.26$      & ~~~~    $3.1^{+1.0}_{-0.9}$  
         \\  
                         $B^{+} \to \bar{D}^{\ast0}D_{s}^{\ast+} $ & ~~~~ $17.1\pm 2.4$&                 ~~0.83~~ &  $B^{+} \to \bar{D}^{\ast0}D_{s1}^{+}(2460) $ & ~~~~ $4.36$ & ~~~~ $3.07$     & ~~~~      $12.0\pm3.0$
         \\  
  \hline \hline
\end{tabular}
\end{table}

 {
 
  \begin{table}[!h]
 \centering
 \caption{Decay constants of $D_{s0}^*(2317)$ and $D_{s1}(2460)$ as the excited states (in units of MeV). \label{Tab:FormFactor1} }
 \begin{tabular}{cccccccc}
 \hline\hline
 Decay Constants ~~~  & $f_{D_{s0}^*}$&   $f_{D_{s1}}$    \\
 \hline
 QCD sum rule~\cite{Wang:2015mxa}~~~  &$333\pm 20$~~~  & $345\pm 17$~~~  \\
 Quark model~\cite{Veseli:1996yg}~~~  &$110$~~~    & $233$~~~  \\
Salpeter method~\cite{Wang:2007av}~~~   &$112$~~~   & $219$~~~ \\ 
covariant light-front quark model~\cite{Verma:2011yw}~~~   &$74.4^{+10.4}_{-10.6}$~~~   & $159^{+36}_{-32}$~~~  \\ 
Lattice QCD~\cite{Bali:2017pdv}~~~   &$114(2)(0)(+5)(10)$~~~   & $194(3)(4)(+5)(10)$~~~  \\ 
Ours~~~   &$115.71$~~~   & $265.74$~~~  \\ 
\hline\hline
 \end{tabular}
 \end{table}

 With the experimental branching fractions of the decays of $B^{+}\to \bar{D}^0 D_{s0}^{*+}(2317)$ and  $B^{+}\to \bar{D}^0 D_{s1}^{*+}(2460)$, we obtain the decay constants $f_{D_{s0}^*}=75.83$~MeV and $f_{D_{s1}}=199.75$~MeV,  corresponding  to physical  $D_{s0}^*$ and $D_{s1}$  as mixtures of molecular and $c\bar{s}$ components.   
 With the obtained $f_{D_{s0}^*}^{M}=58.74$~MeV and $f_{D_{s1}}^{M}=133.76$~MeV in the molecular picture  as well as  the proportions of the molecular components in the total wave functions, e.g., $70\%$ and  $50\%$~\cite{Yang:2021tvc,Song:2022yvz},  one can obtain the decay constants $f_{D_{s0}^*}^{B}=115.71$~MeV and $f_{D_{s1}}^{B}=265.74$~MeV, which correspond to  the picture where $D_{s0}^*$ and $D_{s1}$ are pure excited $c\bar{s}$ states. Table~\ref{Tab:FormFactor1} shows the decay constants of $D_{s0}^*$ and $D_{s1}$ as pure $c\bar{s}$ excited states calculated by several approaches, which are consistent with our estimations and further support the picture where $D_{s0}^*$ and $D_{s1}$ are mainly hadronic molecules but contain sizable $c\bar{s}$ components.  As a result, it is understandable that the branching fractions of the decays  $B \to \bar{D}^{(*)}D_{s0}^*(D_{s1})$ in our calculations are lower than the experimental data.

}       
 
\begin{table}[!h]
\centering
\caption{ { Branching fractions  ($10^{-3}$) of $B_s \to \bar{D}_s^{(\ast)}D_{s0}(D_{s1})$. }\label{resultsbsto2317}
}
\begin{tabular}{c c c | c c c c c c c}
  \hline \hline
   decay modes    &~~~~ Exp~\cite{ParticleDataGroup:2022pth}  &  ~~~ $a_1$~~~    &  decay modes    &~~~~ $\mathrm{Ours}$   
         \\ \hline 
      $B_{s}^{0} \to D_{s}^{+}D_{s}^{-}$    & ~~~~ $4.4\pm 0.5$    & ~~~0.87~~~      &           $B_{s}^{0} \to D_{s}^- D_{s0}^{\ast+}(2317)$    & ~~~~ $0.47$  
        \\
     $B_{s}^{0} \to D_{s}^{\ast+}D_{s}^{-}+ D_{s}^{+}D_{s}^{\ast-}$& ~~~~ $13.9\pm 1.7$ &    ~~~0.83~~~   &              $B_s^0 \to D_s^{*-}D_{s0}^{*+}(2317) $ & ~~~~ $0.27$     
         \\
      $B_{s}^{0} \to D_{s}^{\ast+}D_{s}^{-}+ D_{s}^{+}D_{s}^{\ast-}$& ~~~~ $13.9\pm 1.7$    & ~~~0.77~~~   &             $B_s^0 \to {D}_s^{-}D_{s1}^{+}(2460) $ & ~~~~ $1.18$        \\
      $B_{s}^{0} \to D_{s}^{\ast+}D_{s}^{\ast-}$ & ~~~~ $14.4\pm 2.1$     & ~~~0.84~~~      &      $B_s^0 \to D_s^{\ast-}D_{s1}^{+}(2460) $ & ~~~~ $4.11$    
         \\  
  \hline \hline
\end{tabular}
\end{table}

Along this line, we investigate the decays of $B_s \to \bar{D}_s^{(*)} D_{s}^{(*)}$ and  $B_s \to \bar{D}_s^{(\ast)}D_{s0}(D_{s1})$, which are related to the decays of $B\to \bar{D}^{(\ast)}D_{s}^{(\ast)}$ and $B\to \bar{D}^{(\ast)}D_{s0}^*(D_{s1})$  via   SU(3)-flavor symmetry  as shown in Fig.~\ref{quarktopigy}. The amplitudes for the decays of $B_s \to \bar{D}_s^{(*)} D_{s}^{(*)}$ and $B_s \to \bar{D}_s^{(\ast)}D_{s0}(D_{s1})$ are  the same as those of  their SU(3) symmetric partners. The unknown parameters in the form factors of the $B_s \to \bar{D}_s^{(*)}$  transitions are taken from Ref.~\cite{Verma:2011yw}, tabulated in Table~\ref{BtoDformfactor}. Following the same strategy,  we calculate the branching fractions of the decays $B_s\to \bar{D}_s^{(\ast)}D_{s0}^*$ and $B_s\to \bar{D}_s^{(\ast)}D_{s1}$. The results are shown in Table~\ref{resultsbsto2317}. One can see that the branching fractions of  $D_{s0}^*(2317)$ and $D_{s1}(2460)$ in the $B_s$  decays are similar to those in the $B$ decays, following the  SU(3)-flavor symmetry. Such large production rates { mean }  that the $D_{s0}^*(2317)$ and $D_{s1}(2460)$ are likely to be detected in future experiments.

In addition to the productions of $D_{s0}^*(2317)$ and $D_{s1}(2460)$ in the $B_{(s)}$ decays, it is interesting to investigate their productions in the $\Lambda_b$($\Xi_b$) decays. As indicated in Fig.~\ref{quarktopigy},  the decays of  $\Lambda_b\to D_{s}^{(\ast)}\Lambda_{c}$ and  $\Lambda_b\to D_{s0}^*(D_{s1})\Lambda_{c}$  share the same mechanism as those of $B \to D_{s}^{(\ast)} \bar{D}^{(*)}$ and $B \to D_{s0}^*(D_{s1}) \bar{D}^{(*)}$ at quark level, which   { proceed }  via the decays $b \to c \bar{c}s$.  In terms of  SU(3)-flavor symmetry, we also investigate  the decays  of $\Xi_b\to D_{s}^{(\ast)}\Xi_{c}$ and  $\Xi_b\to D_{s0}^*(D_{s1})\Xi_{c}$.  With the naive factorization approach,  the amplitudes for these decays are given by the effective Lagrangian shown in Eq.~(\ref{Eq:weakdecayV1}), where the parameters in the form factors of the $\Lambda_b \to \Lambda_c$ and $\Xi_b \to \Xi_c$ transitions are obtained in the quark model~\cite{Gutsche:2015mxa,Faustov:2018ahb,Lu:2021irg}, tabulated in Table~\ref{BtoKformfactor1123}. The decay constants of the charmed-strange mesons $D_{s}^{(*)}$ and  $D_{s0}^{*}(D_{s1})$ are calculated in the same way as explained above.

\begin{table}[ttt]
 \centering
 \caption{Values of  $F(0)$, $a$, $b$ in the $\Lambda_b \rightarrow \Lambda_c$ and $ \Xi_b  \rightarrow \Xi_c$  transition  form factors~\cite{Gutsche:2015mxa,Faustov:2018ahb}. \label{BtoKformfactor1123} }
 \begin{tabular}{c|cccccc}
 \hline\hline
   & $F_1^V$~~~ & $F_2^V$~~~ & $F_3^V$~~~ & $F_1^A$~~~ &  $F_2^A$~~~ & $F_3^A$~~~\\
 \hline
 $F(0)^{\Lambda_b \to \Lambda_c}$~~~ &  0.549~~~  & 0.110~~~ & $-0.023$~~~ & 0.542~~~ & 0.018~~~ & $-0.123$~~~\\
 $a^{\Lambda_b \to \Lambda_c}$~~~ &  1.459~~~  & 1.680~~~ & 1.181~~~ & 1.443~~~ & 0.921~~~ & 1.714~~~\\
 $b^{\Lambda_b \to \Lambda_c}$~~~ & 0.571~~~  & 0.794~~~ & 0.276~~~ &0.559~~~ & 0.255~~~ & 0.828~~~\\
\hline 
$F(0)^{\Xi_b \to \Xi_c}$~~~ &  0.467~~~  & 0.145~~~ & $0.086$~~~ & 0.447~~~ & $-0.035$~~~ & $-0.278$~~~\\
 $a^{\Xi_b \to \Xi_c}$~~~ & 1.702~~~  & 2.530~~~ & 1.742~~~ & 1.759~~~ & 2.675~~~ & 2.270~~~\\
 $b^{\Xi_b \to \Xi_c}$~~~ &0.531~~~  & 1.581~~~ & 0.758~~~ &0.356~~~ & 1.789~~~ &1.072~~~\\
\hline\hline
 \end{tabular}
 \end{table}

 \begin{table}[!h]
\centering
\caption{Branching fractions ($10^{-3}$) of    $\Lambda_b\to \Lambda_c D_{s0}^*(D_{s1})$ and $\Xi_b\to \Xi_c D_{s0}^*(D_{s1})$. \label{resultslambdab}
}
\begin{tabular}{c c | c c c c c c c}
  \hline \hline
   decay modes    &~~~~ Exp~\cite{ParticleDataGroup:2022pth,LHCb:2023eeb} &~~~ $a_1$  &    decay modes    &~~~~Ours 
         \\ \hline 
      $\Lambda_{b} \to \Lambda_{c}D_{s}$    & ~~~~ $11\pm 1.0$   &  ~~~$0.88$  & ~~$\Lambda_{b} \to \Lambda_{c}D_{s0}^*(2317)$    & ~~~~ $0.70$  
        \\
      $\Lambda_{b} \to \Lambda_{c}D_{s}^{\ast}$ & ~~~~ $18.568\pm1.102$  & ~~~$0.76$    &        ~~$\Lambda_{b} \to \Lambda_{c}D_{s1}(2460)$   & ~~~~ $4.34$     \\  \hline
   decay modes    &~~~~   Ours  &  ~~~$a_1$    & decay modes    &~~~~ Ours  \\ \hline
               $\Xi_{b} \to \Xi_{c}D_{s}$    & ~~~~ $8.52$  & ~~~$0.88$  &    ~~$\Xi_{b} \to \Xi_{c}D_{s0}^*(2317)$    & ~~~~ $0.58$  
        \\
      $\Xi_{b} \to \Xi_{c}D_{s}^{\ast}$ & ~~~~ $16.30$   & ~~~$0.76$   &       ~~$\Xi_{b} \to \Xi_{c}D_{s1}(2460)$   & ~~~~ $4.29$ 
         \\
  \hline \hline
\end{tabular}
\end{table}

One should note that only the branching fraction of the decay $\Lambda_b\to D_{s}\Lambda_{c}$ is available in the RPP. Very recently   the ratio of $\mathcal{B}(\Lambda_b\to D_{s}^*\Lambda_{c})/\mathcal{B}(\Lambda_b\to D_{s}\Lambda_{c})=1.688\pm 0.022^{+0.061}_{-0.055}$ is reported  by the LHCb Collaboration~\cite{LHCb:2023eeb}, and one can obtain the branching fraction of the decay  $\Lambda_b\to D_{s}^*\Lambda_{c}$.      
With the  branching fractions of $\mathcal{B}(\Lambda_b\to D_{s}\Lambda_{c})$  and $\mathcal{B}(\Lambda_b\to D_{s}^*\Lambda_{c})$ as inputs, we determine $a_1$, and then predict the branching fractions of the decays of
$\Lambda_b\to D_{s0}^*\Lambda_{c}$ and $\Lambda_b\to D_{s1}\Lambda_{c}$, which are shown in Table~\ref{resultslambdab}. We can see that the production rates of   $  D_{s0}^*(2317)$ and $D_{s1}(2460)$   in the $\Lambda_b$ decay are of the order of  $10^{-3}$, which are  
large enough to be detected in future experiments. Since the effective Wilson coefficients in the $B$ decays and $B_s$ decays are similar, as shown in  Table~\ref{resultsbtodds} and Table~\ref{resultsbsto2317}, we can take the same values for $a_1$ in the $\Lambda_b$ decays and the $\Xi_b$ decays. Similarly, we predict the branching fractions of the decays $\Xi_b\to D_{s}^{(\ast)}\Xi_{c}$ and  $\Xi_b\to D_{s0}^*(D_{s1})\Xi_{c}$  in Table~\ref{resultslambdab}. The production rates of ground-state mesons $D_s^{(*)}$ and excited mesons $D_{s0}^*(D_{s1})$ in the $\Xi_b$ decays are of the order of $10^{-2}$ and $10^{-3}$, which are likely to be detected in future experiments.   
In Ref.~\cite{Datta:2003yk}, the authors estimated the ratios  $\mathcal{B}(\Lambda_b \to \Lambda_c M)/\mathcal{B}(B \to D M)$, where $M$ represents the ground-state $D_{s}^{(*)}$ and excited $D_{s0}^*(D_{s1})$ mesons. In Table~\ref{resultsratio}, we show  the  ratios $R_{u}^{M}=\mathcal{B}(\Lambda_b \to \Lambda_c M)/\mathcal{B}(B \to D M)$ and $R_{s}^{M}=\mathcal{B}(\Xi_b \to \Xi_c M)/\mathcal{B}(B_s \to D_s M)$ obtained in this work, which are consistent with Ref.~\cite{Datta:2003yk}. We can see that the production rates of ground-state mesons $D_s^{(*)}$ and excited mesons $D_{s0}^*(D_{s1})$ in the $\Lambda_b(\Xi_b)$   decays are  larger than those in  the  $B_{(s)}$    decays because the  $\Lambda_b \to \Lambda_c$($\Xi_b \to \Xi_c$) form factors are  larger than the corresponding $B_{(s)} \to D_{(s)}$ form factor as shown in Ref.~\cite{Datta:2003yk}.

 \begin{table}[!h]
\centering
\caption{ Ratios of $R_{u}^{M}=\mathcal{B}(\Lambda_b \to \Lambda_c M)/\mathcal{B}(B \to D M)$ and $R_{s}^{M}=\mathcal{B}(\Xi_b \to \Xi_c M)/\mathcal{B}(B_s \to D_s M)$. \label{resultsratio}
}
\begin{tabular}{c c  c | c c c c c c}
  \hline \hline
   $R_{u}^{M}$    &~~~~ Ours &~~~ Ref.~\cite{Datta:2003yk}  &    $R_{s}^{M}$    &~~~~ Ours
         \\ \hline 
      $R_u^{D_{s}}$    & ~~~~ $1.22$   &  ~~~$1.75$   & $R_s^{D_{s}}$     &~~~ 1.94 
        \\
      $R_{u}^{D_{s}^{\ast}}$ & ~~~~ $2.44$  & ~~~$3.47$ & $R_{s}^{D_{s}^{\ast}}$     &~~~ $2.35$      \\  
           $R_u^{D_{s0}^{\ast}}$ & ~~~~ $1.46$  & ~~~$1.58$  & $R_s^{D_{s0}^{\ast}}$     &~~~ $1.23$   \\  
                      $R_{u}^{D_{s1}}$ & ~~~~ $3.12$  & ~~~$4.76$  &  $R_{s}^{D_{s1}}$    &~~~ $3.64$  \\  
  \hline \hline
\end{tabular}
\end{table}

\section{Summary and outlook}
\label{sum}

In this work, we utilized the effective Lagrangian approach to compute the decay constants of $D_{s0}^{\ast}(2317)$ and $D_{s1}(2460)$ as hadronic molecules dynamically generated by the $DK-D_s\eta$ and $D^*K-D_s^*\eta$   contact-range potentials, and then with the naive factorization approach systematically investigated the productions of   $D_{s0}^{\ast}(2317)$ and $D_{s1}(2460)$  in the $B_{(s)}$ and  $\Lambda_b(\Xi_b)$   decays, which proceed via the decay $b \to c \bar{c}s$ at quark level. The decay constants of $D_{s0}^{\ast}(2317)$ and $D_{s1}(2460)$   are estimated to be $58.74$~MeV and $133.76$~MeV. In particular, the decay constant of $D_{s0}^{\ast}(2317)$ is almost independent of the renormalization scale $\mu$ in the loop functions.

As for the branching fractions  of the decays   $B \to \bar{D}^{(*)}D_{s0}^{\ast}(2317)$ and   $B \to \bar{D}^{(*)}D_{s1}(2460)$, our results are smaller than the experimental data, but are consistent with our previous results obtained in the triangle mechanism,   
which {indicate} that  $D_{s0}^{\ast}(2317)$ and $D_{s1}(2460)$ may contain components other than hadronic molecules, such as the $c\bar{s}$ cores.  The values of  $\mathcal{B}[B_s \to  \bar{D}_s^{(*)}D_{s0}^{\ast}(2317)]$ and  $\mathcal{B}[B_s \to   \bar{D}_s^{(*)}D_{s1}(2460)]$ are similar to those of $\mathcal{B}[B \to \bar{D}^{(*)}D_{s0}^{\ast}(2317)]$ and   $\mathcal{B}[B \to \bar{D}^{(*)}D_{s1}(2460)]$, which { reflect}   the underlying SU(3)-flavor symmetry. 
In addition, we predicted the branching fractions of the decays    $\Lambda_b \to  \Lambda_c D_{s0}^{\ast}(2317)$ and  $\Lambda_b \to  \Lambda_c D_{s1}(2460)$ as well as  $\Xi_b \to  \Xi_c D_{s0}^{\ast}(2317)$ and $\Xi_b \to   \Xi_c D_{s1}(2460)$, which are much larger than  the corresponding ones in the $B$ and $B_s$ decays and { indicate  }  that the productions of $D_{s0}^{\ast}(2317)$ and $D_{s1}(2460)$ in the decays of bottom baryons are likely to be detected in future experiments.  

Our study shows that, because of the  equivalence of the triangle mechanism to the tree diagram established in calculating the branching fractions of the decays   $B \to \bar{D}^{(*)}D_{s0}^{\ast}(2317)$ and  $B \to \bar{D}^{(*)}D_{s1}(2460)$, one can extract the decay constants of $D_{s0}^{\ast}(2317)$ and $D_{s1}(2460)$ as hadronic molecules  via the triangle mechanism. This provides an effective approach calculating the decay constants of hadronic molecules, which can then be used in studies of these hadronic molecules in other related processes. We hope that our present work can stimulate more studies along this line.

\section{Acknowledgments}
We are grateful to  Prof. Fu-Sheng Yu  for stimulating discussions.  This work is supported in part by the National Natural Science Foundation of China under Grants No.11975041 and No.11961141004. M.Z.L acknowledges support from the National Natural Science Foundation of China under Grant No.12105007. X.Z.L  acknowledges support from the National Natural Science Foundation of
China under Grant No. 12247159 and  China Postdoctoral
Science Foundation under Grant No. 2022M723149.

\bibliography{reference}
\end{document}